\documentclass[prd,aps,floatfix,nofootinbib,preprint
,tightenlines
]{revtex4}
\usepackage{dcolumn}% Align table columns on decimal point
\usepackage{amsmath}
\usepackage{latexsym}
\usepackage{graphicx}
\usepackage{bm}
\def\svev#1{\left\langle #1\right\rangle}       % variable < >

\def\Tr{{\rm Tr}\,}

\def\Re{{\rm Re\,}}
\def\Im{{\rm Im\,}}

\long \def \blockcomment #1\endcomment{}

\newcommand{\bee}{\begin{equation}}
\newcommand{\ee}{\end{equation}}
\newcommand{\beea}{\begin{eqnarray}}
\newcommand{\eea}{\end{eqnarray}}

\begin{document}
%%%%%%%%%%%%%%%%%%%%%%%%%%%%%%%%%%%%%%%%%%%%%%%%%%%%%%%%%%%%%%%%%%%%%%
\title{%
Volume scaling of Dirac eigenvalues in SU(3) lattice gauge theory with color sextet fermions}
\author{Thomas DeGrand}
\email{degrand@pizero.colorado.edu}
\affiliation{Department of Physics,
University of Colorado, Boulder, CO 80309, USA}
\begin{abstract}
I observe a rough volume-dependent scaling of the low eigenvalues of a chiral Dirac operator
 in lattice studies of SU(3) lattice gauge theory with two flavors of
 color sextet fermions, in its weak-coupling phase.
The mean value of the $i$th eigenvalue scales  with the simulation volume $V=L^4$
as $\langle \lambda_i\rangle L^p \sim \zeta_i$, where $\zeta_i$ is a volume-independent
 constant. The exponent $p$ is about 1.4.
 A possible explanation for this phenomenon is that $p$ is the
  leading relevant exponent associated with the 
fermion mass dependence of correlation functions in a theory whose zero-mass limit is
conformal.
\end{abstract}

%\pacs{11.15.Ha,  12.60.Nz}
%\keywords{Suggested keywords}
\maketitle

%%%%%%%%%%%%%%%%%%%%%%%%%%%%%%%%%%%%%%%%%%%%%%%%%%%%%%%%%%%%%%%%%%%%%
\section{Introduction}
%%%%%%%%%%%%%%%%%%%%%%%%%%%%%%%%%%%%%%%%%%%%%%%%%%%%%%%%%%%%%%%%%%%%%
Recently, various research 
groups
\cite{Damgaard:1997ut,Heller:1997vh,
Catterall:2007yx,Appelquist:2007hu,Shamir:2008pb,Deuzeman:2008sc,DelDebbio:2008zf,Catterall:2008qk,
Fodor:2008hm,DelDebbio:2008tv,DeGrand:2008kx,
Hietanen:2008mr,Fleming:2008gy,Appelquist:2009ty,Hietanen:2009az,Deuzeman:2009mh}
have begun to use lattice methods to
 study field theories which might be candidates for strongly coupled beyond - Standard Model
phenomena \cite{Hill:2002ap}.
 These models typically involve gauge fields and a large number of fermion degrees of freedom,
either many flavors of fundamental representation fermions
(where the discussion goes back to Refs.~\cite{Caswell:1974gg,Banks:1981nn})
 or a smaller number of flavors of 
higher-dimensional representation fermions
(strongly emphasized by Refs.~\cite{Sannino:2004qp,Dietrich:2006cm,Ryttov:2007sr}).

\begin{figure}
\begin{center}
\includegraphics[width=0.7\textwidth,clip]{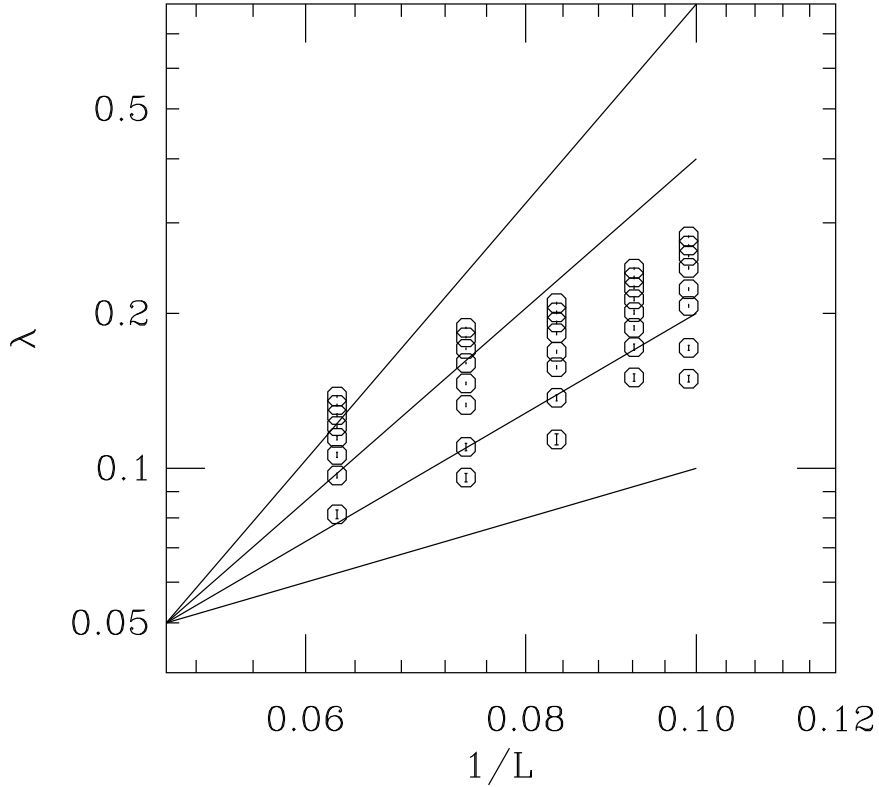}
\end{center}
\caption{Average values of 8 lowest eigenvalues of the sextet-representation
valence overlap operator, vs $1/L$,
where the lattice volume $V$ is defined to be $V=L^4$. The actual volumes, moving
from left to right across the graph, are $16^4$,
$16^3\times 8$, $12^4$, $12^3\times 8$ and $12^3\times 6$.
The four lines are scaling curves $\langle \lambda_i\rangle L^p=$ constant, for $p=4$, 3, 2, 1,
from the top of the figure down.
}
\label{fig:eigvsl}
\end{figure}

 I have been part of a collaboration studying
SU(3) lattice gauge theory with two flavors of
 color sextet fermions\cite{Shamir:2008pb,DeGrand:2008kx}. 
This theory has a weak-coupling phase which is deconfined and chirally restored.
Recently, I observed a peculiar volume-dependent scaling of the eigenvalues of a valence quark
Dirac operator, in the weak coupling phase.
Fig.~\ref{fig:eigvsl} shows this result.
This is very different from the spectrum in simulations with fundamental representation fermions.
 A presentation of the dependence, a comparison with simulations
of two flavors of fundamental representation fermions (which do not show this scaling dependence),
and a possible explanation, are the subjects of this paper.

I begin with some (possibly) relevant background:
The usual description of renormalization for a gauge theory coupled to massless fermions
generally concentrates on the behavior of the running gauge coupling, and makes a distinction about how
the gauge coupling flows under rescaling to the infrared: it could flow
to strong coupling in the case of a confining theory, or to zero, for a trivial theory, or
 to a fixed point at some nonzero value, $g^{2*}$. 
 The latter case is referred to as an
infrared-attractive fixed point (IRFP) theory.
This is a phase with no confinement,
 no chiral symmetry breaking, and algebraic decay of correlation functions (no particles).
Evidence has been presented that several such theories exist:
$SU(3)$ gauge theory with $N_f=12$ flavors of fundamental fermions
 \cite{Appelquist:2007hu,Appelquist:2009ty},
$SU(3)$ gauge theory with $N_f=2$ flavors of sextet fermions (the subject of this paper)
\cite{Shamir:2008pb}
and $SU(2)$ gauge theory with $N_f=2$ flavors of adjoint fermions\cite{Hietanen:2009az}.

While for confining theories the gauge coupling is relevant (the Gaussian fixed point $g=0$ is
unstable), that is not the case for IRFP theories.
The gauge coupling (more precisely the distance of the bare gauge coupling from its fixed point value
$(g^2-g^{2*})$) is an irrelevant coupling. The critical surface encompasses a wide range
of values of bare gauge couplings, plus the values of any other irrelevant operators
(in a lattice theory, these are the usual lattice artifacts).
The relevant coupling of an IRFP theory is the fermion mass, which must be fine-tuned at the
 UV scale to reach
the critical surface (i.e. $m_q=0$). How a nonzero fermion mass
drives the lattice theory off the critical surface determines the asymptotic behavior of (nearly) all
correlation functions.
The situation is completely equivalent to that of an order-disorder transition in a magnetic
 system. The only difference is that the relevant direction
is parameterized by the quark mass $m_q$, rather than the reduced temperature
 $t=(T-T_c)/T_c$ of the magnet. (A closer analogy is to a system which has been fine
 tuned to its Curie point
and then placed in an external magnetic field. The external field breaks the underlying global 
symmetry just as a quark mass explicitly breaks chiral symmetry.) \cite{DeGrand:2009mt}

The framework to describe the physics of these systems is also standard.
Tuning the mass to zero causes the correlation length to diverge algebraically,
\bee
\xi \sim m_q^{-\frac{1}{y_m}}.
\label{eq:corrlen}
\ee
As $m$ is tuned to zero, the singular part of the free energy scales as
\bee
f_s(m_q) = m_q^{D/y_m}(A_1 + A_2  m_q^{|y_i|/y_m}).
\label{eq:fs}
\ee
where $D$ is the system's dimensionality (here $D=4$)
 $A_1$ and $A_2$ are non-universal constants and $y_i$ is the 
biggest non-leading exponent.
This is most likely the exponent $y_g$ of the gauge coupling, $g^2-g^{2*}$, which
 can be determined from the beta function
as measured in (for example) Schrodinger functional simulations at $m_q=0$.

How can one measure $y_m$? The most direct way is through the correlation length $\xi$,
through
Eq.~\ref{eq:corrlen}. 
In a finite simulation volume, $\xi$ will obey Eq.~\ref{eq:corrlen} until it grows
to be the order of the size of the lattice, and then will saturate.
To observe Eq.~\ref{eq:corrlen} probably requires large volumes.

Unitarity bounds for
 conformal field theories\cite{Mack:1976pa} constrain the critical exponent of
 the condensate, $\langle \bar \psi \psi \rangle =\Sigma$,
\bee
\Sigma(m_q) = \frac{\partial f_s}{\partial m_q} \sim m_q^\gamma
\ee
to lie in the range $3>\gamma = D/y_m -1>1$.
(Note that for $D=4$, $\gamma$ is defined to
be $3-\gamma_m$ where $\gamma_m$ is the anomalous dimension for the fermion mass.)
For $D=4$ this means that the allowed range of $y_m$ is between 1 (free field behavior,
all exponents except $y_g$ given by dimensional analysis) and 2.
This  make a measurement of $y_m$ through the mass dependence
of the condensate problematic. The condensate has a UV-sensitive piece
proportional
to the quark mass  \cite{Leutwyler:1992yt}, 
and so $\langle \bar \psi \psi \rangle \sim C_1 m_q + C_2 m_q^\gamma + \dots$. This  masks the
$m_q^\gamma$ non-analytic behavior.
This is quite similar to the situation in finite temperature QCD, precisely at $T=T_c$,
where\cite{Karsch:2008ch}
\bee
\Sigma(a,m_q,T) \sim c_1 m_q/a^2 + c_\delta m_q^{1/\delta} + {\rm analytic}.
\label{eq:sqcd}
\ee
It is different from QCD in that for the IRFP theory there are no Goldstone bosons
(which contribute their own non-analytic piece to the condensate, below $T_c$). It is also
different from QCD in that while in QCD, Eq.~\ref{eq:sqcd} applies only at $T_c$,
in an IRFP theory Eq.~\ref{eq:sqcd} gives the behavior of the condensate
throughout the basin of attraction of the IRFP. Worryingly,
the expected result from $D=3$ $O(4)$ universality, $1/\delta=1/0.56=1.78$, implies that the
 QCD number is  smaller than the smallest unitarity bound, meaning that the
QCD susceptibility would
then be more singular than in a conformal theory. However, the QCD susceptibility
 does not seem to have been measured yet \cite{DeTar:2009ef}.

The way I choose to attack $y_m$ through the condensate involves the
Banks-Casher relation\cite{Banks:1979yr}, which relates the condensate to the
 density of eigenvalues $\lambda$
 of the Dirac operator
$\rho(\lambda)$. At nonzero mass it is
\bee
\Sigma(m_q) = - \int \rho(\lambda) d \lambda \frac{2m_q}{\lambda^2+m_q^2}.
\ee
If the theory is conformal, and $\Sigma(m_q) \sim m_q^\gamma$, then $\rho(\lambda)$ also scales
as $\lambda^\gamma$. The search for the effect I saw
was motivated by a
finite-size scaling argument\cite{Akemann:1997wi} which
 relates the scaling for the density $\rho$ to the scaling of
the value of individual eigenvalues. If we consider the average value of the $i$th eigenvalue
of the Dirac operator in a box of volume $V=L^D$, and  if  $\rho(\lambda)\sim \lambda^\gamma$,
then we expect
\bee
\langle \lambda_i \rangle \sim \left(\frac{1}{L}\right)^p
\label{eq:eigscale}
\ee
where the exponent is
\bee
p=\frac{D}{1+\gamma}.
\label{eq:scaling}
\ee
(A quick derivation: $\rho \sim \lambda^\gamma$ means that eigenvalues are uniformly distributed
in an $N=\gamma+1$ dimensional space of volume $V=R^N$,
\bee
\lambda= \frac{\pi}{R}(\sum_{i=1}^N n_i^2)^{1/2} \ \  n_i=1,2,\dots R 
\ee
 so  an eigenvalue scales as 
$\lambda_i \sim 1/R = (1/V)^\frac{1}{N}=(1/V)^\frac{1}{\gamma+1}$.
 Now suppose we are in $D$ physical dimensions;
in a box of volume $V$, there are  $V=L^D$ modes, from which Eq.~\ref{eq:scaling} is obtained.
One example of this formula is free field theory: $\gamma=D-1$ and $p=1$. Another
is the case of chiral symmetry breaking encoded in the usual formulas of its Random Matrix Theory
analog: $\gamma=0$ so $\rho(\lambda) \rightarrow \rho_0$ a constant, and $p=D$. This is
 an eigenvalue
spectrum which depends on the dimensionless product $\lambda \Sigma V$ or 
$\langle \lambda_i\rangle \sim 1/V$.)

For the case of an IRFP theory, $p$ is equal to $y_m$, the leading 
exponent. A way to determine
$y_m$ is through the volume dependence of eigenvalues of the Dirac operator.

There are several problems associated with carrying out this proposal. The first is numerical.
 Lattice discretizations of the Dirac operator which are easy to implement (allowing
simulations on large volume) break the full $SU(N_f)\otimes SU(N_f)$ chiral symmetry
of the continuum Dirac operator. This complicates the analysis. For example, for
 the fermions in my simulations (Wilson-type fermions), explicit chiral symmetry breaking in the action
introduces an additive shift to the eigenvalues. For a cleaner test of Eq.~\ref{eq:eigscale},
I will use partial quenching: I will take configurations generated
with lattice fermions which do not have exact chiral symmetry, and measure the Dirac spectrum using
an implementation of lattice fermions with exact chiral symmetry (overlap fermions).
This raises another issue: is what the valence fermion sees a faithful realization of what
is happening in the equilibrium distribution of real dynamical variables? I think that for a
first study, what I am going to do is adequate.

The next problem involves interpreting the results. For the case of a system which exhibits
chiral symmetry breaking, there is a tight theoretical description of the behavior of the lowest
eigenvalues of the Dirac operator, which allows one to relate delicate features of the
 spectrum to
the low energy constants of the theory (the condensate, the pseudoscalar decay constant, 
and possibly others).
This description is based on Random Matrix Theory (RMT).  It is quite straightforward to use
Random Matrix Theory predictions to tell whether a system exhibits chiral symmetry breaking.
Conversely,  if a system is
 in a chirally-restored phase, there are no longer RMT predictions to compare results against.

 I already know that our target theory has a weak-coupling phase
which is deconfined and chirally-restored. The restoration of chiral symmetry is observed
through regularities in the spectrum of screening masses as well as the behavior of the
 pseudoscalar
 decay 
constant as a function of quark mass. I can only continue while using the simplest
 properties of the eigenvalues, namely their scaling with system size.

Now I return to a fuller description of Fig.~\ref{fig:eigvsl}.
This is a plot of the average value of the $i$th eigenvalue of the valence overlap operator
versus simulation volume, plotted using $L=V^{1/4}$. The data are all collected at
 the same
 values of the bare parameters,
in the deconfined phase and at a quark mass where the lattice volume causes the 
correlation length to saturate.
 The actual volumes, moving
from left to right across the graph, are $16^4$,
 $16^3\times 8$, $12^4$, $12^3\times 8$ and $12^3\times 6$.
Clearly, the spectrum does not appear to depend on much beside 
the volume, and it seems to show scaling,
$\langle \lambda_i\rangle \sim L^{-p}$.

The four lines are scaling curves $\langle \lambda_i\rangle L^p=$ constant, for $p=4$, 3, 2, 1,
from the top of the figure down. $p=4$ would be the scaling curve if chiral symmetry were broken.
$p=1$ is free-field behavior. The data appear to lie in between -- in fact, the best fit value is 
about $p=1.4$.

Svetitsky, Shamir and I have already observed (but with a different
 lattice discretization) that $SU(3)$ $N_f=2$ sextet QCD
shows evidence for an IRFP \cite{Shamir:2008pb}.
 If that observation survives future tests, the exponent $p$ is indeed $y_m$, the leading
relevant exponent. If that observation turns out to be false, the volume scaling of eigenvalues is still
something striking. It is different from what is seen in the deconfined phase
 $SU(3)$ $N_f=2$ fundamental QCD (as we will see, below).

The data in Fig.~\ref{fig:eigvsl} all come from simulations at one set of bare parameters.
 However, if the simulations were
in fact done in the basin of attraction of the IRFP,
irrelevancy of the gauge coupling means
 that different gauge couplings merely correspond to lattice actions with
different amounts of scaling violations. 

Finally, the idea behind this calculation was
 first described by the authors of Ref.~\cite{Fodor:2008hm}. They proposed doing
simulations with dynamical
 chiral fermions (overlap fermions) so that the Dirac operator whose eigenvalues
are measured is the same one that appeared in the action. They performed these simulations,
also in the deconfined phase of  $SU(3)$ $N_f=2$ sextet QCD.
 They identified  that the eigenvalue spectrum was inconsistent
 with expectations for the behavior of a system with chiral symmetry breaking,
but only studied  one volume ($6^4$ lattices).

 The paper proceeds as follows. In Sec.~\ref{sec:two} I describe details of the simulations.
In Sec.~\ref{sec:three} I provide some background: I examine whether lattice data show evidence
that the gauge coupling is irrelevant.
I make a direct attack on the correlation length exponent  using Eq.~\ref{eq:corrlen}.
I find numbers also in the range 1.5-1.6.
 I then make some comparisons of $N_f=2$ fundamental
and sextet QCDs in their deconfined phases. These systems are quite different.
In Sec.~\ref{sec:four} I provide background for Fig.~\ref{fig:eigvsl} and describe my
attempts
to pin down the exponent $p$.
I conclude with some speculations and (rather obvious) suggestions for follow-up work.

\section{Numerical techniques\label{sec:two}}
I  performed simulations on a system with $SU(3)$ gauge fields and two flavors of 
dynamical fermions
in the symmetric (sextet) representation of the color gauge group.
The lattice action is defined by the single-plaquette gauge action and a
Wilson fermion action with added clover term~\cite{Sheikholeslami:1985ij}.
The fermion action employs
the differentiable hypercubic smeared link of
 Ref.~\cite{Hasenfratz:2007rf}, from which the symmetric-representation gauge connection
for the fermion operator is constructed. No tadpole-improvement is used and the
clover coefficient is set to
its tree-level value.
The bare parameters which are inputs to the simulation are the gauge 
coupling $\beta=6/g^2$, the fermion hopping parameter $\kappa$.
 The integration is done with  one additional heavy
pseudo-fermion field as suggested by Hasenbusch \cite{Hasenbusch:2001ne},
multiple time scales \cite{Urbach:2005ji},
and a second-order Omelyan integrator \cite{Takaishi:2005tz}.

The routines for simulating sextet-representation fermions were developed
with (and mostly by) B.~Svetitsky and Y.~Shamir. The dynamical fermion algorithm was adapted
from a program written by A.~Hasenfratz, R.~Hoffmann and S.~Schaefer\cite{Hasenfratz:2008ce}.
All computer code is based on the publicly available code of the MILC collaboration~\cite{MILC}.

Simulation volumes range from $12^3 \times 6$ to $16^4$ sites, 
and typical data sets range from
a few hundred to a thousand trajectories. I recorded lattices every five trajectories 
(of unit simulation time trajectory length) and collected 40-80 lattices per parameter set
for the calculation of screening lengths and  overlap eigenvalues.

The trick of combining periodic and anti-periodic boundary conditions for valence quarks
\cite{Blum:2001xb,Aoki:2005ga,Allton:2007hx,Aubin:2007pt}
is used in spectroscopy or screening mass measurements.

Throughout this work, instead of quoting $\kappa$, I will use the the Axial Ward Identity (AWI) quark mass,
 defined through
\bee
\partial_t \sum_x \svev{A_0(x,t)X(0)} = 2m_q \sum_x \svev{ P(x,t)X(0)}.
\label{eq:AWI}
\ee
where $A_0=\bar \psi \gamma_0\gamma_5 \psi$, $P = \bar \psi \gamma_5 \psi$, and $X$ is any source.
(I temporarily drop factors of the lattice spacing in the derivations.)
The derivative is taken to be the naive  difference
operator ($\partial_\mu f(x)=(f(x+\hat\mu a) - f(x-\hat\mu a))/(2a)$).

The valence Dirac operator whose eigenvalues are the subject of this study is the overlap
operator \cite{Neuberger:1997fp,Neuberger:1998my}.
 Details of the particular implementation of the action are described in
 Refs.~\cite{DeGrand:2000tf,DeGrand:2004nq,DeGrand:2006ws,DeGrand:2007tm,DeGrand:2006nv}.
The only new ingredient is the application to symmetric-representation fermions, using the
same combination of hypercubic link and projection as for the dynamical fermions.
Eigenvalues of the squared Hermitian Dirac operator $D^\dagger D$ are computed
 using the ``Primme'' package of
McCombs and Stathopoulos\cite{primme} and split apart in the usual way.

There are potential problems with this analysis.  The first one involves the
index theorem, relating the winding number of the gauge field $k$
to the number of Dirac fermion zero modes,
\bee
{\rm index} = 2T(R) k.
\ee
 $T(R)$ is the Dynkin index of the representation $R$, 1/2 for fundamental representation fermions,
 $(N+2)/2$ for sextet fermions in the color group $SU(N)$, and so on.
 Thus we expect to see
multiples of 5  zero modes for adjoint overlap fermions in our $SU(3)$ case.

However, ten years ago Heller, Edwards, and Narayanan discovered\cite{Heller:1998cf}
 that the index theorem applied
to adjoint overlap fermions in background $SU(2)$ gauge configurations failed: here $2T(R)=4$ and they saw
configurations with zero modes which were not multiples of four. Similar results were more
 recently reported by
Garcia Perez, Gonzalez-Arroyo and Sastre\cite{GarciaPerez:2007ne}.
N\'ogr\'adi reported similar behavior from simulations with $SU(3)$ sextet fermions\cite{Nogradi}
 and in simulations where the bare gauge coupling is large, I have seen
configurations whose zero mode content was a not a multiple of 5.

It is unknown whether this is a disease, just a particular failure of the overlap action to capture
topology when the gauge configuration is rough, or real physics, something which persists in a
 continuum limit.
I am going to ignore it. The gauge configurations
at the parameter value used to construct  Fig.~\ref{fig:eigvsl} were smooth enough that
all  of the lattices I collected for Fig.~\ref{fig:eigvsl} had $Q=0$.

The second potential problem concerns possible phase structure in the theory.
In Ref.~\cite{DeGrand:2008kx} we observed that in the deconfined phase of sextet QCD, the Polyakov
 loop ordered
in one of the negative real directions, roughly along one of the complex elements of $Z(3)$
 ($\Re \langle  \Tr P(x)\rangle <0$, $\Im \langle  \Tr P(x) \rangle \ne 0$).
 This is quite different behavior from
fundamental QCD, where the Polyakov loop orders positively  ($\Re \langle  \Tr P(x) \rangle >0$,
$\Im \langle  \Tr P(x)  \rangle=0$). It is likely that there
is a complicated phase structure as a function of the bare parameters, with different vacua
 favored in different regions of the bare parameter space \cite{Myers:2009df}.
Detailed studies of the phase structure (at smaller volumes) are currently being performed by
Machtey, Shamir, and Svetitsky \cite{MSS}. For the time being, I believe, based on my simulations,
 that the equilibrium
 small quark mass region is $\Re\langle \Tr P(x) \rangle <0$.
I checked that all the data in  Fig.~\ref{fig:eigvsl} remained in the  $\Re \langle \Tr P(x) \rangle <0$,
$\Im \langle \Tr P(x)\rangle \ne 0$, phase during
 their time histories.

\section{Some spectroscopic checks\label{sec:three}}

Before analyzing eigenvalues, I pause to consider some potential complications.

\subsection{Does the gauge coupling appear to be irrelevant?}
If the gauge coupling is irrelevant, observables collected at different values of the bare coupling
will show small differences, qualitatively similar to the different sizes of scaling violations seen
in ordinary QCD simulations when different lattice actions are used.

We \cite{DeGrand:2008kx} saw this behavior in our earlier work, but did not describe it as such. It is easy to overlay
screening masses taken in the deconfined phase at different values of $\beta$
 (at identical lattice volumes)
and observe that they coincide. I now show a variation of that comparison.

\begin{figure}
\begin{center}
\includegraphics[width=0.6\textwidth,clip]{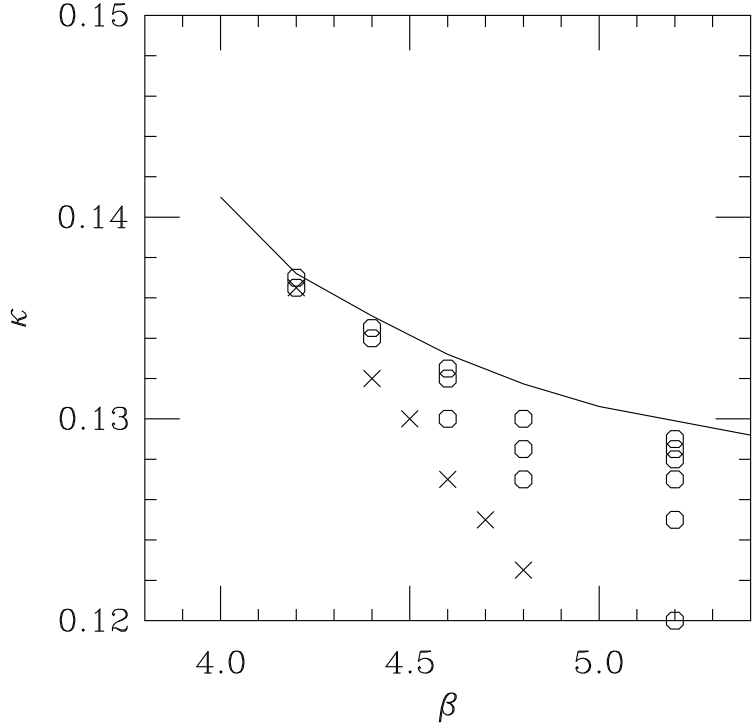}
\end{center}
\caption{Map of the bare coupling constant plane relevant to our sextet simulations. The solid line
is the line of zero quark mass, $\kappa=\kappa_c$. The crosses show the location of the
 confinement-deconfinement crossover at $N_t=6$. Octagons show the data collection points
for the next figures.
\label{fig:betakappa}}
\end{figure}

To begin, the phase diagram of the present system is shown in Fig.~\ref{fig:betakappa}. The line is the
location where the AWI quark mass vanishes, $\kappa=\kappa_c$. The crosses show the location of the
$N_t=6$ deconfinement transition. To the left of the crosses,
 the theory confines and (to the extent we can discuss this for massive quarks) chiral symmetry
 is broken; to the right, the theory is deconfined and chiral symmetry is restored. Octagons show
 the data collection points (all on $12^3\times 6$ volumes)
for the next figures. I have checked that all data collected at larger volumes (all at $\beta=5.2$)
is deconfined, but have not accurately mapped the phase boundary.

\begin{figure}
\begin{center}
\includegraphics[width=0.5\textwidth,clip]{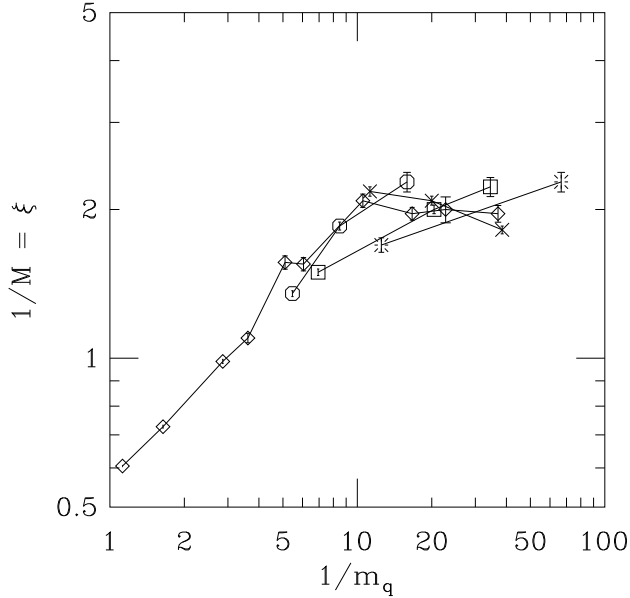}
\end{center}
\caption{Correlation length (inverse pseudoscalar screening mass) vs inverse quark mass
for $N_f=2$ sextet fermions on a $12^3\times 6$ volume.
Bursts are $\beta=4.2$; crosses, $\beta=4.4$; squares, $\beta=4.6$, octagons, $\beta=4.8$, and diamonds,
$\beta=5.2$. Lines connect points with the same $\beta$ values.
\label{fig:xivsm126}}
\end{figure}

I define the correlation length $\xi$ to be the inverse of the
screening mass in the pseudoscalar channel
and plot $\xi$ vs $1/m_q$, the inverse AWI fermion mass, in Fig.~\ref{fig:xivsm126}.
(In the deconfined phase, the pseudoscalar, vector, axial vector and scalar screening masses are
essentially degenerate.)
The data  show
that the correlation length is dominantly driven by $1/m_q$, much less by the gauge coupling $\beta$.

\subsection{Contrasting the deconfined phase for $N_f=2$ fundamental fermions}

\begin{figure}
\begin{center}
\includegraphics[width=0.5\textwidth,clip]{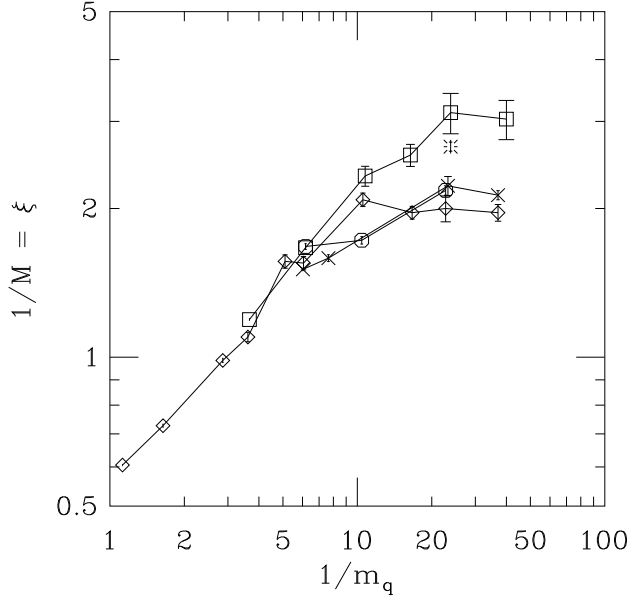}
\end{center}
\caption{Correlation length (inverse pseudoscalar screening mass) vs inverse quark mass
for $N_f=2$ sextet fermions at $\beta=5.2$.
Diamonds,  $12^3\times 6$ volume;
octagons,  $12^3\times 8$ volume.
crosses,  $12^4$ volume;
squares,  $16^3\times 8$ volume,
and burst, $16^4$ volume.
 Lines connect points with the same $\beta$ values.
\label{fig:xivsmN}}
\end{figure}

The reader might recall a fact about screening masses in the deconfined phase of fundamental
representation QCD: they scale with  system size like \cite{screening}
\bee
m_H^2 = 4\left [ \left(\frac{\pi}{N_t}\right)^2 + m_q^2\right]
\label{eq:minmat}
\ee
where $\frac{\pi}{N_t}$ is the lowest nonzero Matsubara frequency associated with
antiperiodic boundary conditions in a lattice of temporal length $N_t$. 
The data in Fig.~\ref{fig:xivsm126} are for $N_t=6$,
for which Eq.~\ref{eq:minmat} predicts $m_H =  2\pi/6 \sim 1$ in the chiral limit. The
limiting value of $m_H$ is about half that value.  In fact, Eq.~\ref{eq:minmat} does not reproduce
any of the data. The situation at the one $\beta$ value (5.2) where I have many volumes,
ranging from $12^3\times 6$ to $16^4$, is shown in
 Fig.~\ref{fig:xivsmN}. 
The limiting correlation length does not seem to depend separately on $N_t$; rather, it appears to
increase as the lattice volume increases.

\subsection{Scaling of correlation lengths?}
In the infinite volume limit we expect that
 the mass gap will disappear at vanishing fermion mass.
The chiral susceptibility $\chi = \partial \Sigma(m_q)/\partial m_q$ is related to the volume
integral of the isosinglet correlator $C(x)=\langle \bar \psi \psi(x) \bar \psi \psi(0)\rangle$.
When the lightest mass in this channel vanishes,  the susceptibility
will diverge. As in QCD, this argues that the screening mass in this channel should vanish.
Again, as in QCD, the calculation of isosinglet correlators is difficult due to the presence
of disconnected diagrams, so as a first try I will mimic the QCD analysis \cite{DeTar:2008qi}
 and look at the isotriplet
scalar correlator. Just because the data is quite accurate, 
I will also look in the pseudoscalar channel.

Fig.~\ref{fig:xivsmN} shows us that data from different volumes seem to lie on a volume independent
 curve until the correlation length reaches a size comparable to the system size. Presumably the
relation between rounding of the correlation length and volume is geometry dependent.
So I will work with the two data sets I have, which have the same aspect ratio, 
$12^3\times 6$ and $16^3\times 8$. I will try a fit of the form of Eq.~\ref{eq:corrlen},
varying the range and composition.

\begin{figure}
\begin{center}
\includegraphics[width=0.8\textwidth,clip]{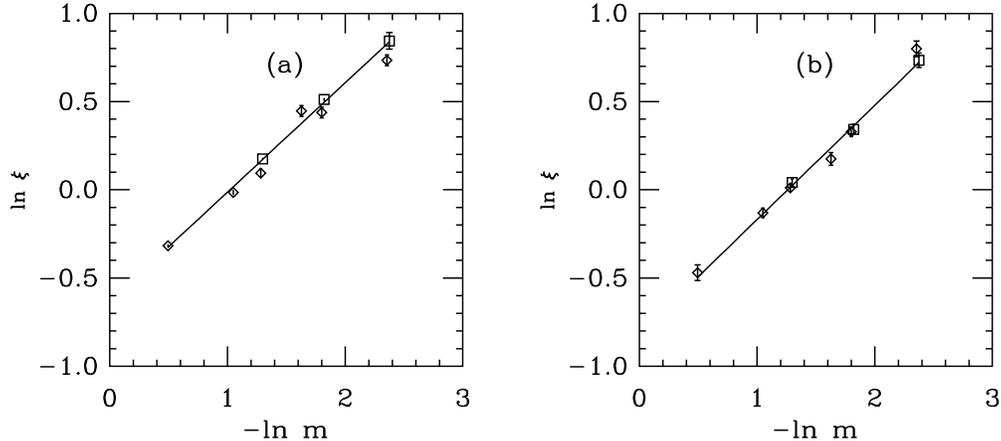}
\end{center}
\caption{Representative fits to (a) pseudoscalar and (b) scalar correlation lengths.
Data are diamonds for $12^3\times 6$ volumes and squares for $16^3\times 8$ volumes.}
\label{fig:corrfit}
\end{figure}

A pair of representative fits are shown in Fig.~\ref{fig:corrfit}. The left panel is the pseudoscalar
and the right panel the scalar; I will define an exponent through $1/m_j \sim (1/m_q)^{1/y_j}$
with $j=\pi$ or $a_0$.
 For the pictured fits the results are 
%$1/y_m=0.60(2)$,
% (or 
$y_\pi=1.66(6)$,
 $\chi^2=41.7$/7 dof
(pseudoscalar) and 
%$1/y_m=0.65(2)$ (or 
$y_{a_0}=1.54(5)$,
 $\chi^2=9.7$/7 dof (scalar). It is relatively easy to
find ranges of fits for the scalar mesons for which $\chi^2<10$ with $1/y_{a_0}$ drifting 
from .51(2) to .67(3)
according to the fit range. The pseudoscalar data has smaller uncertainty and hence poorer
confidence levels,  $\chi^2$/dof between 3-4 for 4-9 points, and $1/y$ ranging from 0.57(1) to 0.62(2).
My data do not have the dynamic range to take my result too seriously.
(If people had not spent the last 30 years fitting $m_\pi^2 \sim m_q$ in QCD over the same range
 of masses, I would not have tried it.) Researchers with larger simulation
volumes might like to attempt  this, however.

\section{Discussion of results from eigenvalues\label{sec:four}}
\subsection{Volume scaling?}
Since I have argued that the gauge coupling appears to be irrelevant, results from any gauge coupling
anywhere in the deconfined phase (in the basin of attraction of the FP?) should be equally meaningful.
I chose a gauge coupling $\beta=5.2$ and $\kappa=0.1285$.
These are the second-smallest masses in Fig.~\ref{fig:xivsmN}.
At this gauge coupling simulations are easy to perform. At this hopping parameter the IR cutoff
(lattice volume) dominates the effect of the quark mass (the AWI mass is about 0.042).
I computed the values of the lowest 8 eigenvalues of the Hermitian squared Dirac operator $D^\dagger D$,
which I then converted into eigenvalues of the overlap operator by the usual stereographic projection.
All of the configurations collected at this parameter value have zero topological charge.
Results were shown in  Fig.~\ref{fig:eigvsl}.

\begin{figure}
\begin{center}
\includegraphics[width=0.8\textwidth,clip]{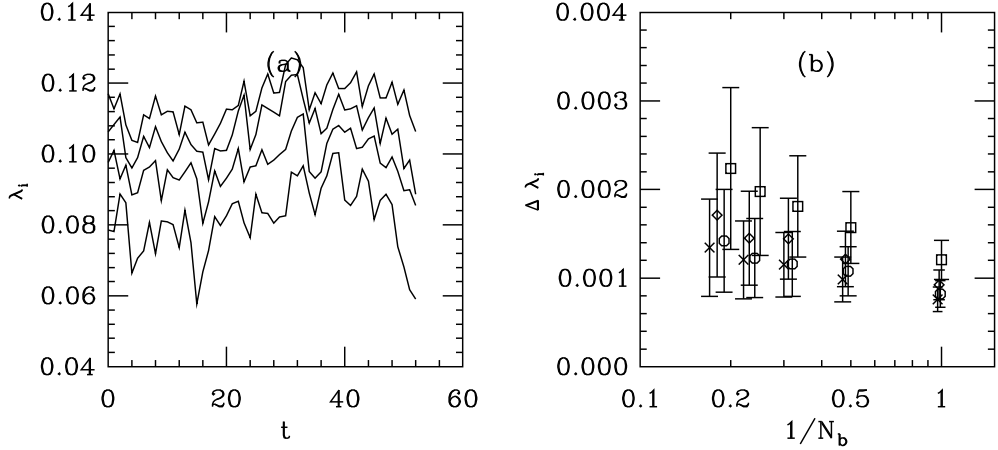}
\end{center}
\caption{(a) Time history of overlap eigenvalues from the $16^4$ data set (in units of 5 HMC time steps).
(b) Uncertainty on the average $\langle \lambda_i \rangle$ as a function of bin size.
Symbols are squares for $i=1$,
octagons for $i=2$,
diamonds for $i=3$,
and crosses for $i=4$.}
\label{fig:example}
\end{figure}

Two pictures illustrate the quality of the data, combined into Fig~\ref{fig:example}. The left panel shows 
simulation time histories of the lowest four eigenvalues of the $16^4$ data set.
Each measurement is separated by five HMC trajectories. The right panel
shows the error on the average computed by blocking $N_b$ successive measurements together.
The lowest eigenvalue is clearly the noisiest, but the autocorrelation time seems not to be too large:
I bin two successive lattices ($N_b=2$ or $\Delta t = 10$ HMC units) together before
averaging.

The spectrum is nothing like the spectrum of eigenvalues from a theory with chiral symmetry breaking,
which consists of roughly equally spaced energy levels extending from the origin. The large spectral gap
separating the lowest eigenvalue from the origin reflects the absence of chiral symmetry breaking
in the weak coupling phase.

\begin{figure}
\begin{center}
\includegraphics[width=0.5\textwidth,clip]{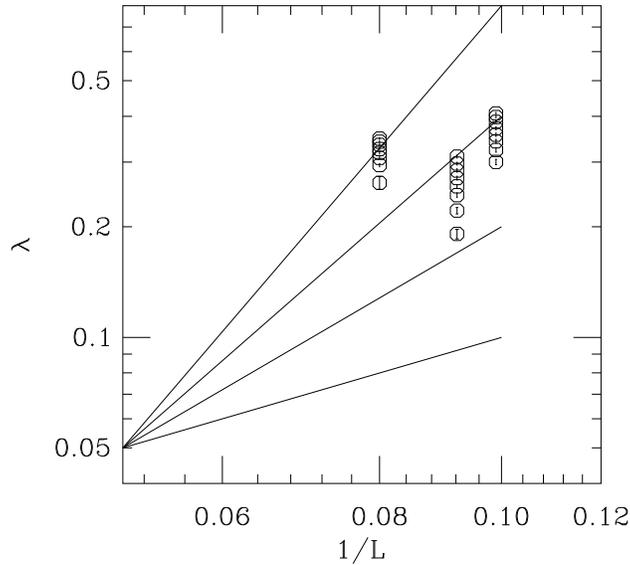}
\end{center}
\caption{Average value of $i$th eigenvalue of the fundamental-representation
valence overlap operator, vs $1/L$,
where the lattice volume $V$ is defined to be $V=L^4$, from fundamental 
representation simulations.
 The actual volumes, moving
from left to right across the graph, are $16^3\times 6$,  $12^3\times 8$ and $12^3\times 6$.
The four lines are scaling curves $\langle \lambda_i\rangle L^p=$ constant, for $p=4$, 3, 2, 1,
from the top of the figure down.
}
\label{fig:eigvslF}
\end{figure}

I repeated the calculation of eigenvalues using
fundamental-representation valence quarks from a fundamental-representation simulation
in the deconfined phase. The gauge coupling is $\beta=5.5$ and $\kappa=0.126$. This is 
a parameter set in which the screening masses have saturated at their Matsubara values.
The picture is completely different -- see Fig. \ref{fig:eigvslF}. Again, there is a spectral gap,
but now it is dominantly affected by the short (antiperiodic) length of the lattice,
Matsubara physics again. The eigenvalue distribution does not depend  simply  on $L=V^{1/4}$.

Now let us try to extract an exponent from Fig. \ref{fig:eigvsl}. This is not so easy: 
for theoretical input, all we have is Eq.~\ref{eq:eigscale}, the finite-size scaling formula.
One does not know a priori if it applies to all the eigenvalues or to only the lowest eigenvalues.
One also does not know if there is some minimum volume for which it applies. (All of these
questions have RMT answers for chirally-broken systems, but that is not what we are
 looking at here).
So I will just proceed empirically: I will look at fits to individual eigenvalues, then groups of them.
I will fit all the data sets or drop smaller volumes and fit only the larger ones.

\begin{figure}
\begin{center}
\includegraphics[width=0.7\textwidth,clip]{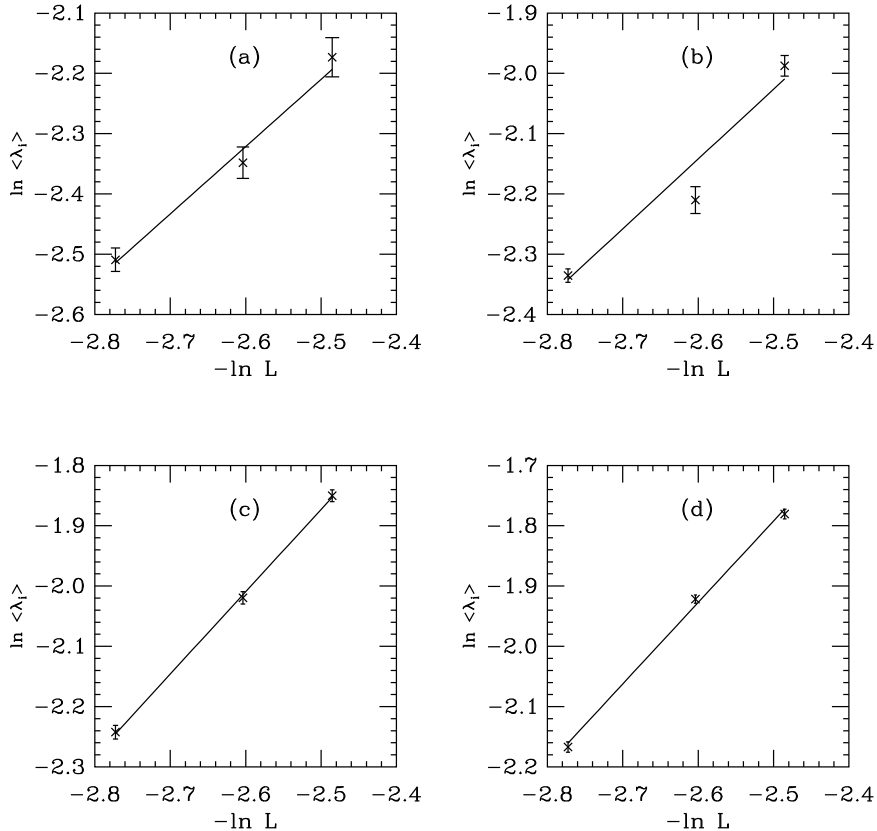}
\end{center}
\caption{Three-volume fits to individual eigenvalues, (a) the lowest eigenvalue
(b) the first excited state (c) the second excited state (d) the third excited state.
The volumes are (from the left) $16^4$, $16^3\times 8$ and $12^4$.
}
\label{fig:several3}
\end{figure}

\begin{figure}
\begin{center}
\includegraphics[width=0.7\textwidth,clip]{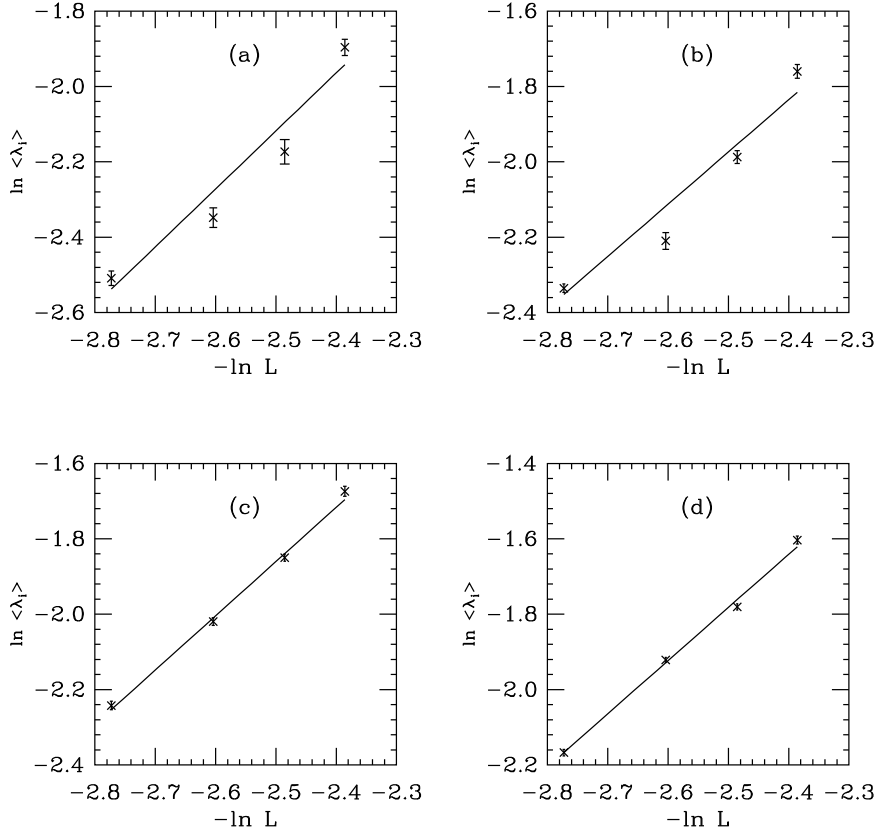}
\end{center}
\caption{Four-volume fits to individual eigenvalues, (a) the lowest eigenvalue
(b) the first excited state (c) the second excited state (d) the third excited state.
The volumes are (from the left) $16^4$, $16^3\times 8$, $12^4$, and $12^3\times 8$.
}
\label{fig:several4}
\end{figure}

I begin by fitting
individual eigenvalues (lowest, second, and so on) to a power law,
$\ln \langle \lambda_i \rangle = A_i - p \ln L$. I choose to fit to all five volumes,
or four, or the largest three. The individual data points in each fit are uncorrelated, of course.
Fits and chi-squareds are shown in Table \ref{tab:fittable}.
Examples of fits are shown in Figs.~\ref{fig:several3} and \ref{fig:several4}.

\begin{table}
\caption{Exponent $p$ from fits to individual eigenvalues, from the largest three 
volumes ($16^4$, $16^3\times 8$, $12^4$), or 
the largest four (add $12^3\times 8)$, or to all volumes (add $12^3\times 6$).}
\begin{center}
\begin{tabular}{c|cc|cc|cc}
\hline\hline
  &  3 volumes & & 4 volumes & &  5 volumes \\
mode &  $p$ &  $\chi^2$ &  $p$ &  $\chi^2$ &  $p$ &  $\chi^2$ \\
\hline
1 & 1.11(12)& 1.2  &  1.54(7) & 19.8 & 1.42(5) & 25.6 \\
2 & 1.16(7) & 10.1 &  1.39(5) & 32.4 & 1.31(4) & 37 \\
3 & 1.36(5) & 0.4   & 1.44(4) & 5.8 & 1.47(3) & 6.9 \\
4 & 1.35(4) & 4.0  & 1.41(3) & 9.3 & 1.44(2) & 11.3 \\
5 & 1.46(4) & 22   & 1.44(3) & 22.7 &1.49(2) & 29 \\
6 & 1.49(3) & 57   & 1.44(2) & 60.5 & 1.48(2) & 67 \\
7 & 1.52(3) & 88   & 1.45(2) & 101 & 1.48(2) & 109 \\
8 & 1.52(3) & 100  & 1.44(2) & 114 & 1.49(2) & 129 \\
\hline\hline
\end{tabular}
\end{center}
\label{tab:fittable}
\end{table}

\begin{figure}
\begin{center}
\includegraphics[width=0.7\textwidth,clip]{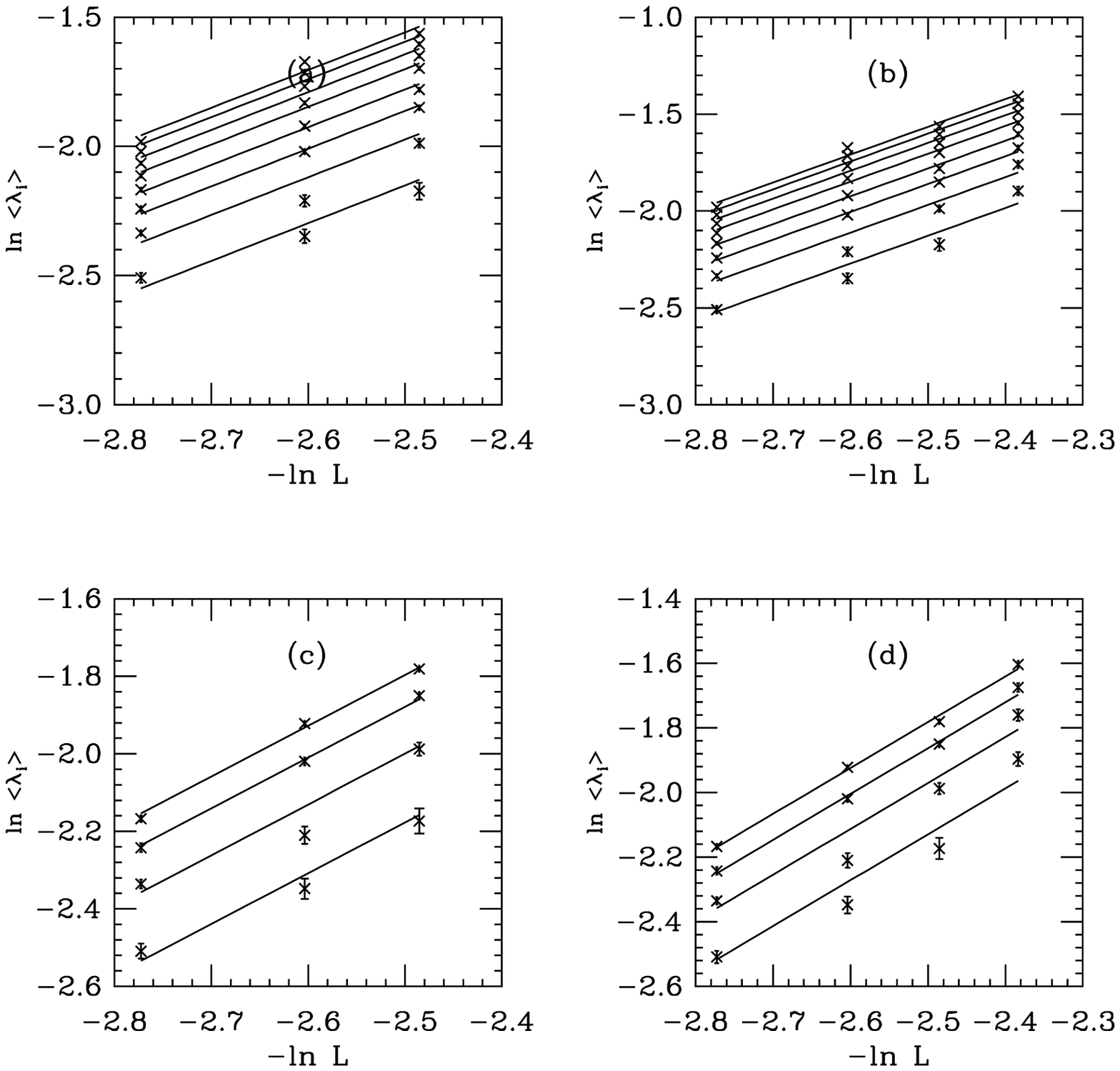}
\end{center}
\caption{Combined fits to several eigenvalues.
(a) Three-volume fits to the 8 lowest eigenvalues, (b) four volume fits to the lowest 8
eigenvalues,
(c) Three-volume fits to the  lowest 4 eigenvalues, (d) four volume fits to the lowest 4
eigenvalues.
The volumes are (from the left) $16^4$, $16^3\times 8$, $12^4$ and in (b) and (d) $12^3\times 8$.
}
\label{fig:several}
\end{figure}

Clearly, I can also fit groups of eigenvalues. In that case, I fit
$\ln \langle \lambda_i \rangle = A_i - p \ln L$ for $i=1\dots N$ eigenvalues.
Now the data are correlated. I look at the quality of fits from uncorrelated
fits, and then repeat by taking bootstrap averages of the data.
The behavior is quite similar to the fits to individual eigenvalues. Some examples
(with bootstrap errors) on the average, chi-squared from uncorrelated fits:
\begin{itemize}
\item Fit the lowest 4 eigenvalues and biggest 3 volumes: $p=1.30(4)$, $\chi^2/dof=25/(12-5)$
\item Fit the lowest 4 eigenvalues biggest 4 volumes: $p=1.42(3)$, $\chi^2/dof=67/(16-5)$
\item Fit all 8 eigenvalues and biggest 3 volumes: $p=1.46(3)$, $\chi^2/dof=331/(24-9)$
\item Fit all 8 eigenvalues and biggest 4  volumes: $p=1.43(2)$, $\chi^2/dof=372/(32-9)$
\end{itemize}
Examples of these fits are shown in Fig. \ref{fig:several}.

Generally, the fits are poor but the trend of the four-volume fits is clear: $p\sim 1.4$.
I cannot assign an error which is not a guess: there are too many unknowns. Are the volumes
 large enough?
Are the data sets large enough? Nobody has tried
such an analysis before. I think it is better to present my results and let others 
explore new directions.
Nevertheless, if the reader desires  a number with an error bar,
 $p=1.4(1)$ will capture the uncertainty over which fit to choose.

 \subsection{Additional properties of eigenvalues}
\begin{figure}
\begin{center}
\includegraphics[width=0.7\textwidth,clip]{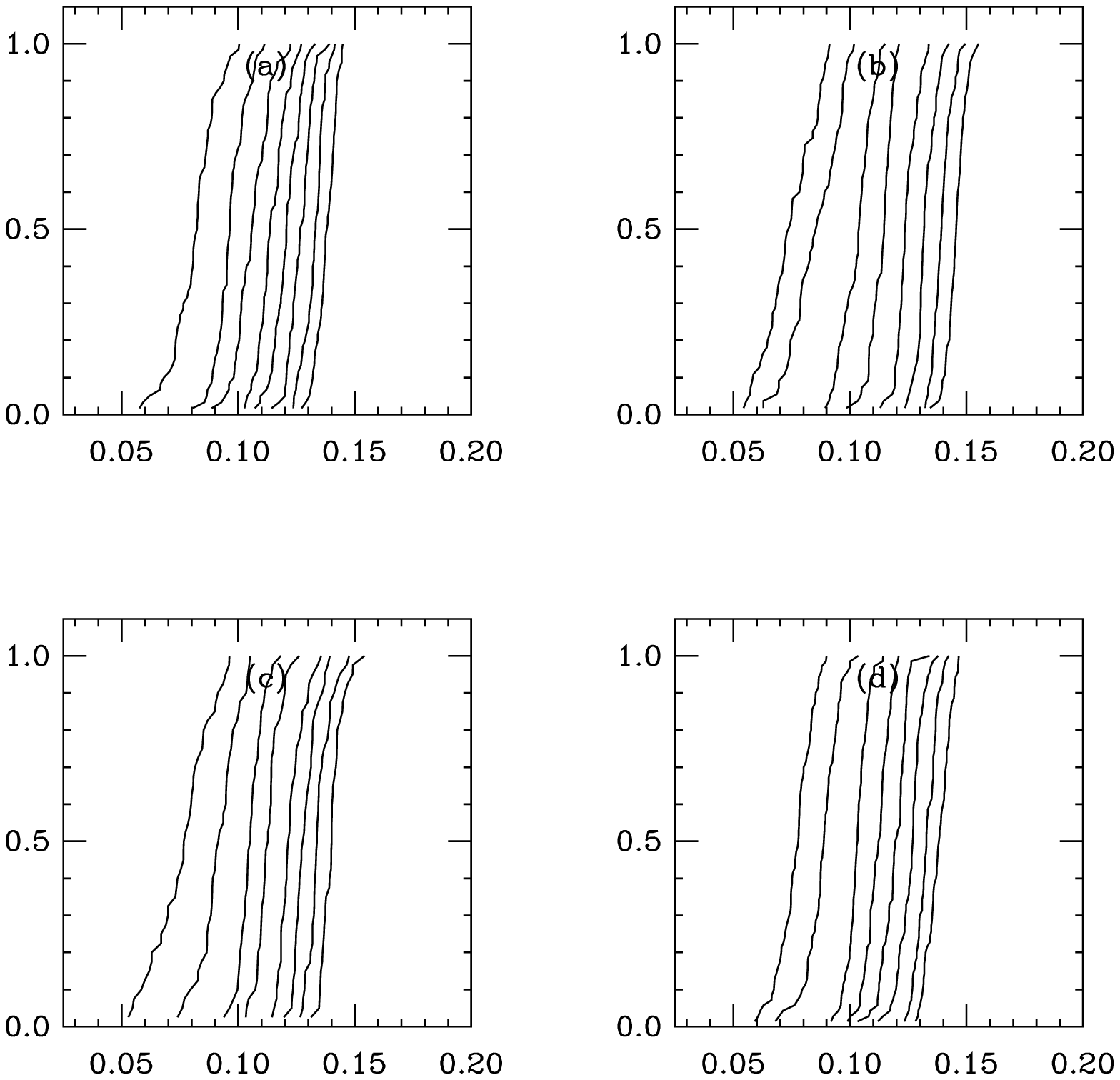}
\end{center}
\caption{Cumulants, where the eigenvalue is rescaled by  $(L_i/L_1)^p$ with $p=1.4$
for the four largest volumes: (a) $16^4$ (no rescaling) (b) $16^3\times 8$ (c) $12^4$
 (d) $12^3\times 8$.
}
\label{fig:cumulant}
\end{figure}

Of course, there is more to the eigenvalue distribution than its average.
In the absence of any theory, I will just compare cumulants,
$C(x) = n(x)/N$ where $n(x)$ is the number of data points with a value smaller
than $x$ and $N$ the total number of data points. I will rescale the data at each volume
by a factor $(L_i/L_1)^p$ with $p=1.4$ and $L_1=16$ in
 the plots. This is shown in Fig~\ref{fig:cumulant}.

Apart from the lowest eigenvalue the cumulants are very sharp, indicating a strongly localized
eigenvalue spectrum. This was observed by the authors of Ref.~\cite{Fodor:2008hm}.
The lowest eigenvalue has a broad distribution. This is not surprising; the spectral gap is large
and the ``pressure'' on it from all the higher eigenvalues is asymmetric. Could the broad
 distribution be related to
the relatively poor scaling fits for this mode?

\begin{figure}
\begin{center}
\includegraphics[width=0.5\textwidth,clip]{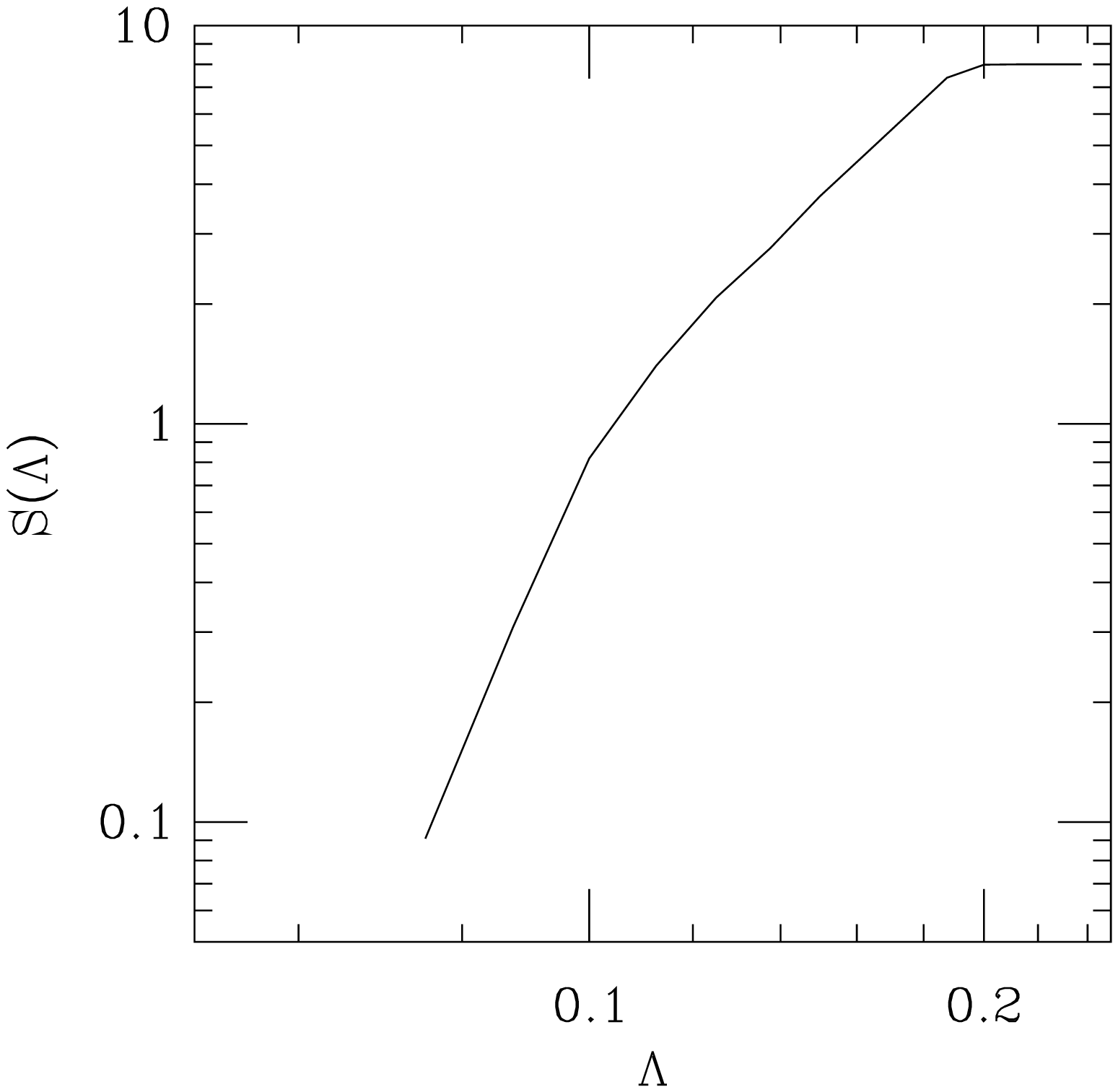}
\end{center}
\caption{$S(\Lambda)$ vs $\Lambda$ (defined in
Eq.~\protect{\ref{eq:nuov}}) from the $16^3\times 8$ data set.
}
\label{fig:nuov}
\end{figure}

A direct calculation of $\rho(\lambda)$ involves histogramming the data. A possibly better
technique is inspired by Ref.~\cite{Giusti:2008vb}, namely to integrate the density
\bee
S(\Lambda) = \int_0^\Lambda \rho(\lambda) d\lambda 
\sim \Lambda^{\gamma+1}.
\label{eq:nuov}
\ee
(basically the cumulant of all eigenvalues less than $\Lambda$).
This was not a successful way to measure $\gamma$, since I did not have enough eigenmodes.
Fig.~\ref{fig:nuov}
illustrates a typical data set, this time the $16^3\times 8$ one.
 A few minutes with a ruler will produce a power law in the 
vicinity of 3 for $1<S(\Lambda)<8$ (compared to $4/y_m=2.9$ for $y_m=1.4$)
 but this is clearly only of entertainment value:
the curve flattens when the data sets are exhausted, and presumably at lower $\Lambda$ values 
the higher (missing) modes would contribute as well. Ref.~\cite{Giusti:2008vb} analyzed far more
eigenvalues.
I can collect several tens of eigenvalues of the squared clover Dirac operator quite easily, but
the analysis did not seem to be crisp due to the explicit chiral symmetry breaking in the action,
shifting the zero,
and I did not pursue this as far as I probably should have.

At low $\Lambda$ the plot shows a clear break. This is the part of the data which is completely
dominated by the lowest eigenvalue, whose low-$\lambda$ edge is not given by a simple power law.
There must be interesting physics here (as there is for eigenvalue distributions of chirally broken
systems) but I do not know what it is.
I also do not know if this is real physics or just statistics: if $\rho(\lambda)$ is small in some
 region of $\lambda$, a big data set is needed to populate it.

\section{Discussion\label{sec:five}}

Cautious readers might conclude that this paper is content-free: a curious Fig. \ref{fig:eigvsl}
which appears only at the end of a long set of improbable  constructions. In fact, I
 present the paper
because  Fig. \ref{fig:eigvsl} is so striking.
Most of my analysis assumed that the underlying theory was conformal in the massless,
infinite
volume limit. I believe this assumption but it needs verification. Whether it is or not,
one fact seems true: The deconfined phase of $SU(3)$ gauge theory with $N_f=2$ flavors of
sextet fermions seems to have quite different properties 
from the  deconfined phase of $SU(3)$ gauge theory with $N_f=2$ flavors
of fundamental fermions.

Let us assume that the exponent I was measuring is truly what I claim it is: the leading
exponent of a theory with one relevant coupling (the mass) which is critical at zero mass.
As far as a lattice simulator can tell from reading the continuum literature, the 
answer $y_m \sim 1.4$ is very interesting: scaling exponents far from their engineering
dimensions seem to be much sought after. (A partial subset of the literature is
Refs.~\cite{Cohen:1988sq,Braun:2006jd,Appelquist:1996dq,Appelquist:1998rb,gardi,Kaplan:2009kr}.)
 Recalling that the exponent of the condensate
is $\gamma =4/y_m -1$, this is an exponent of about $\gamma=1.9(2)$. 

A common prediction in the literature is that a lower bound on $\gamma$
is $\gamma=2$ or $y_m=4/3$. The earliest
appearance of this result I can find is by Cohen and Georgi \cite{Cohen:1988sq},
who obtain it by solving a gap equation.
In a recent paper Kaplan, Lee and Son \cite{Kaplan:2009kr} argue
on different grounds that $\gamma$ should always exceed 2.
My result  is barely consistent with this prediction,
 with an uncertainty which in the end seems a bit inadequate.

Ryttov and Sannino \cite{Ryttov:2007cx}
have a supersymmetric QCD - inspired beta function for gauge theories
with higher dimensional representations of fermions. The anomalous dimension is predicted to be
\bee
\gamma = 3 - \frac{11C_2(G)-4T(R)N_f}{2T(R)N_f} = 3 - \frac{13}{10} = 1.7.
\ee

There are some models of technicolor where the new physics sector is conformal or nearly so 
(for example \cite{Luty:2004ye,Luty:2008vs});
large anomalous dimensions ($\gamma$ near 1) are desirable to achieve their phenomenological goals.
This does not seem to be the case for sextet QCD.

The theoretical literature almost always addresses many models simultaneously,
and validating (or otherwise)
their assumptions requires data from many systems.
$y_m$'s for other lattice IRFP theories are needed.

From a simulation point of view, the present study could be improved by
eliminating partial quenching.
Perhaps the best candidate to do this is $SU(2)$ color with adjoint fermions. Codes may be fast enough
to allow for simulations with dynamical overlap fermions on reasonable volumes.

A limitation of the use of eigenvalues is that at present, the theory is just finite size scaling.
RMT predictions were rich because the number of theoretically motivated observables was large.
Analysis like what I am doing needs additional theoretical input.

The published simulations of $N_f=12$ fundamental fermions use staggered fermions.
The technology for measuring susceptibilities for the condensate in the context of QCD
thermodynamics is quite well developed. It would be interesting to apply it to present needs.
Staggered fermion eigenvalues are also  cheap to compute
but explicit flavor breaking effects could hamper their interpretation.

Ultimately, the only universal quantities associated with IRFP theories are their critical exponents.
Can lattice simulations measure them?

\begin{acknowledgments}
%%%%%%%%%%%%%%%%%%%%%%%%%%%%%%%%%%%%%%%%%%%%%%%%%%%%%%%%%%%%%%%%%%%%%%
I thank P.~Damgaard, C.~DeTar, A.~Hasenfratz,   T.~G.~Kovacs, D.~N\'ogr\'adi, 
B.~Svetitsky and Y.~Shamir for discussions.
This work was supported in part by the US Department of Energy.
%
%%%%%%%%%%%%%%%%%%%%%%%%%%%%%%%%%%%%%%%%%%%%%%%%%%%%%%%%%%%%%%%%%%%%
\end{acknowledgments}
%%%%%%%%%%%%%%%%%%%%%%%%%%%%%%%%%%%%%%%%%%%%%%%%%%%%%%%%%%%%%%%%%%%%%

\end{document}